\documentclass[aps,prc,twocolumn,showpacs,groupedaddress,floatfix]{revtex4}

\usepackage{graphicx}
\usepackage{dcolumn}
\usepackage{amsfonts}

\begin{document}

\title{TDHF fusion calculations for spherical+deformed systems}

\author{A.S. Umar and V.E. Oberacker}
\affiliation{Department of Physics and Astronomy, Vanderbilt University,
             Nashville, Tennessee 37235, USA}

\date{\today}


\begin{abstract}
We outline a formalism to carry out TDHF calculations of fusion cross sections for spherical + deformed
nuclei. The procedure incorporates the dynamic alignment of
the deformed nucleus into the calculation of the fusion cross section.
The alignment results from multiple E2/E4 Coulomb excitation of the ground state rotational band. 
Implications for TDHF fusion calculations are discussed.
TDHF calculations are done in an unrestricted three-dimensional geometry using modern
Skyrme force parametrizations.
\end{abstract}
\maketitle


\section{\label{sec:intro}Introduction}

Heavy-ion fusion reactions are a sensitive probe of the size and
structure of atomic nuclei~\cite{Ba80}. Recent experiments with radioactive ion beams
allow for the study of heavy-ion fusion with exotic nuclei. For
instance, enhanced fusion-evaporation cross sections have been observed with
neutron-rich $^{132}Sn$ beams on $^{64}Ni$~\cite{Li03}. The synthesis of
superheavy nuclei in hot and cold fusion
reactions~\cite{Ho02,Og04,Gi03,Mo04,II05} represents another experimental
frontier.

In general, the fusion cross sections
depend on the interaction potential and form factors in the
vicinity of the Coulomb barrier. Furthermore, experiments on subbarrier
fusion have demonstrated a strong dependence of the total fusion cross section
on nuclear deformation~\cite{SE78}. The dependence on nuclear orientation has received
particular attention for the formation of heavy and superheavy elements~\cite{KH02}
and various entrance channel models have been developed to predict its role in
enhancing or diminishing the probability for fusion~\cite{FG04,SC04}. While this
may be true for heavy systems, the orientation effects should influence the
fusion process for light nuclei as well.

There are several theoretical methods for calculating heavy-ion fusion
cross sections:
a) barrier penetration models~\cite{TB84,RO83,BT98},
b) coupled-channels calculations~\cite{LP84,RP84,Esb04,HR99}, and
c) microscopic many-body approaches such as
   the Time-Dependent Hartree-Fock (TDHF) method~\cite{JN82,BFH85,U91,UO05}.

In recent coupled-channels calculations of heavy-ion fusion reactions, one typically
uses the ``rotating frame approximation''~\cite{Esb04,HR99} which assumes that the
orbital angular momentum $L$ of relative motion is conserved; this approximation
avoids the full angular momentum coupling and thus reduces the number of
coupled channels considerably. One tends to use empirical interaction potentials
which are mostly real and energy-independent (Woods-Saxon potential, proximity-type
potentials, double-folding potential). An imaginary potential is unnecessary
because one takes into account explicitly all channel couplings that affect fusion.
Coupling potentials are usually obtained utilizing
macroscopic nuclear structure models (e.g. rotational model or harmonic vibrator).
The coupled-channel equations are solved numerically with standard scattering
boundary conditions at $r \rightarrow \infty$, and an incoming-wave boundary condition
at some point $r=R_F$ inside the Coulomb barrier. The fusion cross section is obtained
from the incoming flux at $R_F$.

Fusion in the TDHF collision process is achieved when the relative kinetic
energy in the entrance channel is entirely converted into internal
excitations of a single well defined compound nucleus. In the TDHF theory
the dissipation of relative kinetic energy into internal excitations is
due to the collisions of the nucleons with the ``walls'' of the
self-consistent mean-field potential. TDHF studies demonstrate that the
randomization of the single-particle motion occurs through repeated
exchange of nucleons from one nucleus into the other. Consequently, the equilibration of
excitations is very slow and it is sensitive to the details of the
evolution of the shape of the composite system. If one of the nuclei is
deformed, one needs to carry out TDHF fusion calculations on a 3-D lattice 
for different relative orientations of the nuclei as they approach each other on a 
Rutherford trajectory, and afterwords one has to perform a suitable average over
all orientations. Depending on the incident energy and impact parameter, some
relative orientations may contribute to fusion while others may not. In lowest-order
approximation, one might assume that all relative orientations of the intrinsic axis
system occur with equal probability. This is indeed a reasonable approximation for
relatively light nuclei where Coulomb excitation is negligible. However, in heavier
systems, the dynamical alignment of the deformed nucleus as a result of multiple
E2 and E4 Coulomb excitation of the ground state rotational band may no longer be ignored
and results in a definite preferential alignment of the deformed nucleus which must
be taken into account when calculating the fusion cross section.

There are a very limited number of TDHF collision studies involving deformed nuclei,
and almost all of these were done in an axial geometry using the ``rotating frame
approximation'', and thus could not address the orientation dependence of reaction
cross sections.
Recently, we have developed a new TDHF code which assumes no symmetries regarding 
both the collision geometry and the Skyrme effective interaction~\cite{UO06a} and
utilizes the modern Skyrme force parametrizations, including the time-odd terms.
The numerical calculations are performed on a large 3-D Cartesian
lattice using the Basis Spline collocation method~\cite{U91} for increased accuracy.
In this work we shall use this code to study fusion cross sections in the collisions
of spherical+deformed nuclei.

The theoretical formalism
for calculating the fusion cross section is outlined in Section~\ref{sec:fus_theory}.
In Section~\ref{sec:results} we present fusion calculations carried out using this
approach. In Section~\ref{sec:summary} a summary and outlook is provided.

\section{\label{sec:fus_theory}TDHF with alignment}

The evaluation of the heavy-ion collision dynamics can be divided into two separate steps:
a) A dynamical Coulomb alignment calculation to determine the probability that a given nuclear orientation
occurs at the distance $r(t_0)$, where the TDHF run is initialized. The distance $r(t_0)$ is chosen such
that the nuclei only interact via the Coulomb interaction.
b) A TDHF calculation, starting at this finite internuclear distance $r(t_0)$, for
a fixed initial orientation of the deformed nucleus.
\subsection{Coulomb excitation}
\begin{figure}[!htb]
\includegraphics*[scale=0.50]{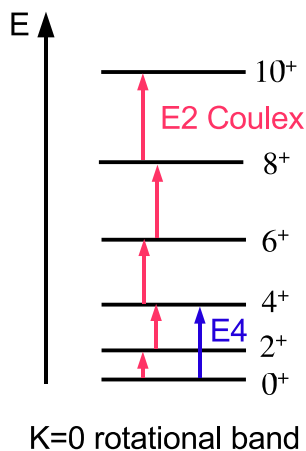}
\caption{\label{fig:rot_band} Coulomb excitation of the ground
state rotational band members via multiple E2 and E4 transitions.}
\end{figure}

\begin{figure}[!htb]
\includegraphics*[scale=0.45]{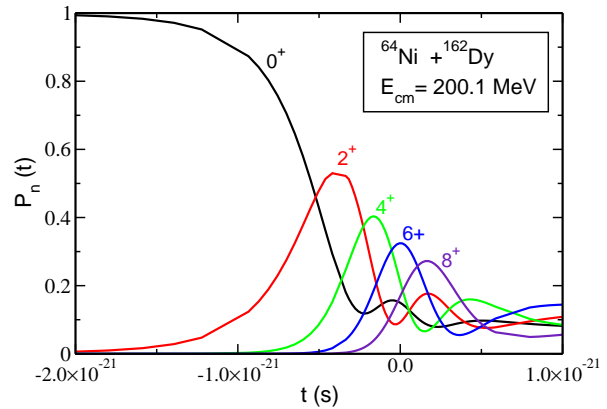}
\caption{\label{fig:prob_of_t} Coulomb excitation probability of the $0^+,...,8^+$ members
of the ground state rotational band in $^{162}Dy$ as a function of time. The deformed
nucleus is excited with a $^{64}Ni$ beam at $E_{cm}=200.1$ MeV at impact parameter $b=0$.}
\end{figure}

\begin{figure}[htb]
\includegraphics*[scale=0.30]{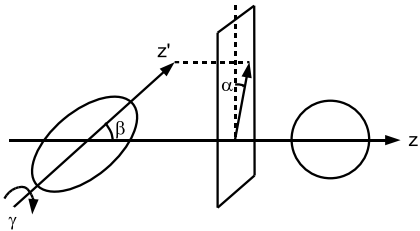}
\caption{\label{fig:align_schem} The orientation of the intrinsic body-fixed
frame of the deformed target nucleus is specified by the three Euler angles $(\alpha,\beta,\gamma)$
relative to the collision plane.}
\end{figure}

For a given incident energy $E_{cm}$ and impact parameter $b$, we carry out a semiclassical
Coulomb excitation calculation of the dominant ground state rotational band of the deformed
nucleus (see Fig.~\ref{fig:rot_band}).
The Coulomb excitation calculation starts at very large internuclear distances
(about $1500$ fm) when both nuclei may be presumed to be in their respective
ground states. We have provided the details of this formalism in the Appendix.
Fig.~\ref{fig:prob_of_t} shows a representative result in medium-mass nuclei,
$^{64}_{28}Ni + ^{162}_{\ 62}Dy$, where we plot the excitation
probability of the $^{162}Dy$ ground state rotational band members as a function of time.
In the Coulomb excitation code, we utilize the measured energy levels of the ground state
rotational band in $^{162}Dy$ up to the $18^+$ level \cite{ENSDF}. We also use experimental values for the reduced
transition probabilities \cite{TOI} $B(E2,0^+ \rightarrow 2^+) = 5.35 e^2 b^2$ and 
$B(E4,0^+ \rightarrow 4^+) = 0.07 e^2 b^4$ from which one can calculate all
remaining E2 and E4 matrix elements within the collective rotor model. 
In this semiclassical calculation, one can see how the nucleus ``climbs up the
rotational band'' during the collision. Negative times in Fig.~\ref{fig:prob_of_t}
correspond to the incoming branch of the Rutherford trajectory, and positive times
correspond to the outgoing branch. The distance of closest approach is reached
at $t=0$. (Alternatively, one can plot the excitation probability as a function
of the internuclear distance). The Coulomb excitation amplitudes, $a_{JM} (t)$,
give rise to a preferential orientation (alignment). The dynamic alignment
formalism presented in the Appendix allows us to follow the nuclear alignment as a
function of the internuclear distance vector ${\bf r}(t)$.
We would like to stress that the quantity which enters in our TDHF fusion calculations
is not the Coulomb alignment after the reaction has taken place (at $t \rightarrow +\infty$)
but rather the alignment at the \emph{finite internuclear distance} $r(t_0) \approx 15 fm$
on the incoming branch of the Rutherford trajectory.
In the case of a spherical+deformed system,
the relative orientation of the deformed nucleus may be specified by the three
Euler angles $(\alpha,\beta,\gamma)$ relative to the collision plane as shown in
Fig.~\ref{fig:align_schem}. We use here the definition of the Euler angles given in
Ref.~\cite{BS79}. Multiple Coulomb
excitation predominantly excites the members of the $K=0$ ground state rotational band
which preserves an axially symmetric shape; hence, rotations about the symmetry axis $z'$
described by the Euler angle $\gamma$ are irrelevant (see Appendix).

\subsection{Fusion cross section}
Subsequent to the determination of the initial conditions for the two nuclei,
as described in the previous section, we perform a TDHF calculation to establish
the outcome of the collision process. In TDHF, for a given set of initial conditions
(energy, impact parameter, orientation, etc.), only one outcome is possible, thus we
define
\begin{equation}
P_{\mathit{TDHF}}(b,E_{\mathit{cm}};\beta ,\alpha )= 
\left\{ \begin{array}{l}
1 \;\;\;\; \textrm{\small TDHF fusion}\\
0 \;\;\;\; \textrm{\small TDHF no fusion},
\end{array}\right.
\end{equation}
where $b$ is the asymptotic impact parameter, $E_{cm}$ is the associated center of
mass energy, and quantities $\beta$ and $\alpha$ denote the orientation angles
discussed above.
The total fusion cross section may be written as
\begin{equation}
\sigma_{fusion}(E_{cm}) = 2\pi\int_0^{b_{max}} b\; db\; P_{fusion}(b,E_{cm})\;.
\label{sig1}
\end{equation}
The probability for fusion is given in terms of the differential probability
\begin{equation}
P_{fusion}(b,E_{cm}) = \int d\Omega \;\frac{dP_{fusion}(b,E_{cm};\beta,\alpha)}{d\Omega}\;,
\label{sig2}
\end{equation}
where the Euler angle solid angle element is given by $d\Omega = sin \beta d\beta d\alpha$, and
\begin{eqnarray}
\frac{dP_{fusion}(b,E_{cm};\beta,\alpha)}{d\Omega} &=& \frac{dP_{orient}(b,E_{cm},t_0;\beta,\alpha)}{d\Omega}
      \nonumber \\
 &\times& P_{\mathit{TDHF}}(b,E_{\mathit{cm}};\beta ,\alpha ) \ .
\label{sig3} 
\end{eqnarray}
Since for a given $E_{cm}$ value and angles $(\beta,\alpha)$ the TDHF fusion probability
is either zero or unity we can instead write the integrals over the impact parameter to
terminate at $b_{max}(\beta,\alpha)$,
where $b_{max}(\beta,\alpha)$ denotes the maximum impact parameter for fusion for orientation
angles $\beta$ and $\alpha$. This allows us to write
\begin{eqnarray}
\frac{d\sigma_{fusion}(E_{cm};\beta,\alpha)}{sin\beta d\beta d\alpha} &=& 2\pi\int_0^{b_{max}(\beta,\alpha)} b\; db \nonumber \\
 &\times& \frac{dP_{orient}(b,E_{cm},t_0;\beta,\alpha)}{sin\beta d\beta d\alpha}\; .
\label{sig4} 
\end{eqnarray}
The differential cross section given in Eq.~(\ref{sig4}) in terms of Euler angles 
should not be confused with the laboratory differential cross section given 
in terms of the scattering angles. The differential orientation probability used in
Eq.~(\ref{sig4}) is evaluated in the Appendix, Eq.~(\ref{eq:orient_prob}).


\section{\label{sec:results}Numerical studies}
In order to gain a better insight into the differential orientation probability mentioned
above we have performed calculations for systems that involve two heavy reaction partners
and for systems which are composed of two light nuclei.
We first examine the dynamical orientation, due to multiple E2/E4 Coulomb excitations,
of the deformed nucleus $^{162}_{\ 66}$Dy colliding with a spherical
$^{64}_{28}$Ni nucleus at $E_{cm}=265$ MeV. In a central collision (impact parameter $b=0$),
the orientation probability of an even-even nucleus
is independent of the Euler angle $\alpha$
because, starting from the $0^+$ ground state, only magnetic substates $M=0$ of the
ground state rotational band are excited. Therefore, the orientation probability depends only
on the Euler angle $\beta$ in this case (see Appendix, Eq.~\ref{eq:orient_prob}).
Fig.~\ref{fig:align_ni_dy} shows the orientation probability for two values of the
internuclear distance. At very large distance ($R=1363$ fm) before the collision
where the deformed nucleus is in its $0^+$ ground state, all Euler angles $\beta$
occur with equal probability, and there is no preferential alignment.
However,
at an internuclear distance of $R=R_1+R_2+2=13.2$ fm (where $R_1$ and $R_2$ 
denote the mean square nuclear radii), we observe substantial alignment preference.
Clearly, the perpendicular orientation ($\beta=90^0$) is preferred over the 
parallel ($\beta=0^0$) orientation, with an alignment ratio
$dP(\beta=90^0)/d\Omega \  / \ dP(\beta=0^0)/d\Omega = 1.54$.
The reason for this preferred alignment  is easily understood in terms of
classical electrostatics: for a given internuclear distance $R$, the 
perpendicular orientation of the deformed nucleus minimizes the Coulomb
interaction energy which is proportional to $P_2(cos \beta)$.
\begin{figure}[htb]
\includegraphics*[scale=0.40]{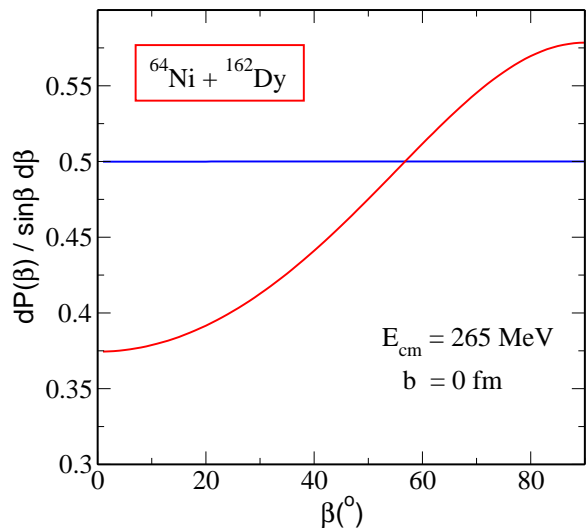}
\caption{\label{fig:align_ni_dy} Dynamic alignment due to Coulomb
excitation of $^{162}$Dy by a $^{64}$Ni beam. Shown is the orientation probability
as a function of the Euler angle $\beta$ in a central collision at internuclear
distances $R=1363$ fm (blue curve) and at $R=13.2$ fm (red curve).}
\end{figure}

We have repeated the above analysis to study the Coulomb excitation of the
deformed nucleus $^{22}_{10}Ne$ in a central collision with an $^{16}_{\ 8}O$ nucleus.
In the Coulomb excitation code, we utilize the measured energy levels of the $^{22}Ne$
rotational band up to the $6^+ (6.311 MeV)$ level \cite{ENSDF}. The 
$B(E2,0^+ \rightarrow 2^+)$ value is computed from the measured halflife \cite{TOI}
$T_{1/2} = 3.63 ps$, and the remaining E2 matrix elements are determined from
the collective rotor model.
\begin{figure}[!htb]
\includegraphics*[scale=0.40]{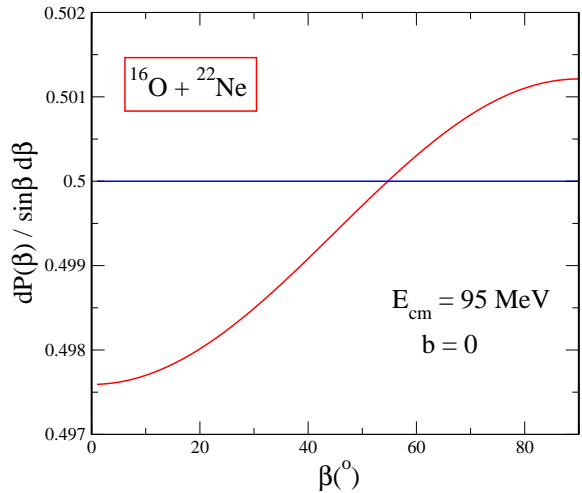}
\caption{\label{fig:align_ox_ne} Dynamic alignment due to Coulomb
excitation of $^{22}$Ne by a $^{16}$O beam. Shown is the orientation probability
as a function of the Euler angle $\beta$ in a central collision at internuclear
distances $R=154$ fm (blue curve) and at $R=14$ fm (red curve).}
\end{figure}
Fig.~\ref{fig:align_ox_ne} shows the corresponding orientation probability
for these two light nuclei. Despite the
much higher incident energy of $E_{cm}=95$ MeV (about $6.9$ times the Coulomb
barrier height), we observe only a very small alignment preference:
$dP(\beta=90^0)/d\Omega \  / \ dP(\beta=0^0)/d\Omega = 1.007$.
There are two aspects pertaining to the importance of dynamical alignment
in calculating fusion cross sections. In the case where all orientations lead to
fusion the alignment probability provides a properly weighted sum for calculating
the fusion cross section using Eq.~(\ref{sig2}). However, implicit in the above
expression is also the fact that for different orientations we will have different
values for $b_{max}$. Thus, even in a case where the central collision leads to
fusion for all orientations of the deformed nucleus, and even if the Coulomb
excitation probabilities are angle-independent, we still have different
contributions to the fusion cross section arising from different orientations.
\begin{figure*}[!htb]
\begin{center}
\includegraphics*[scale=0.32]{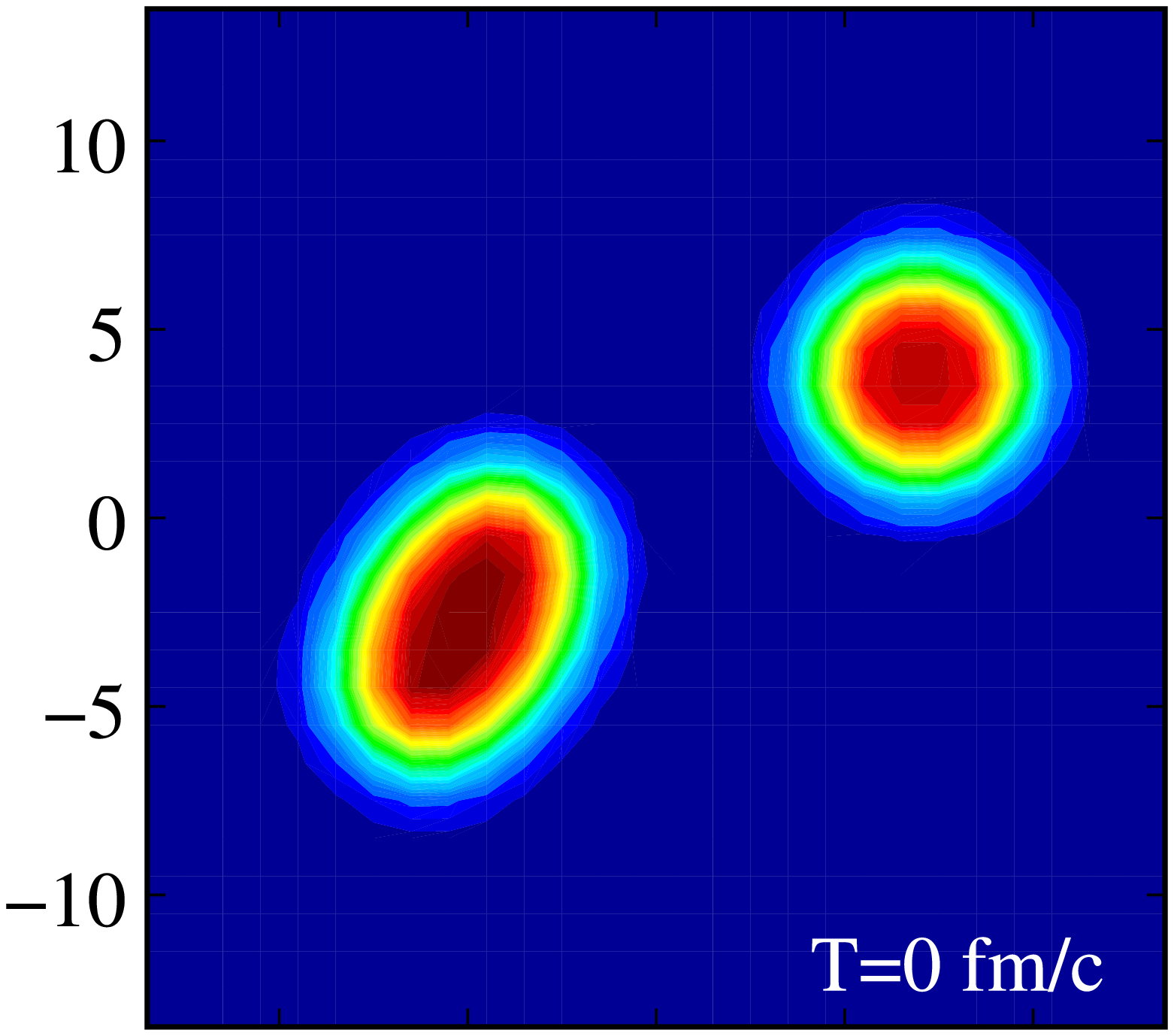}\hspace{-0.05in}
\includegraphics*[scale=0.32]{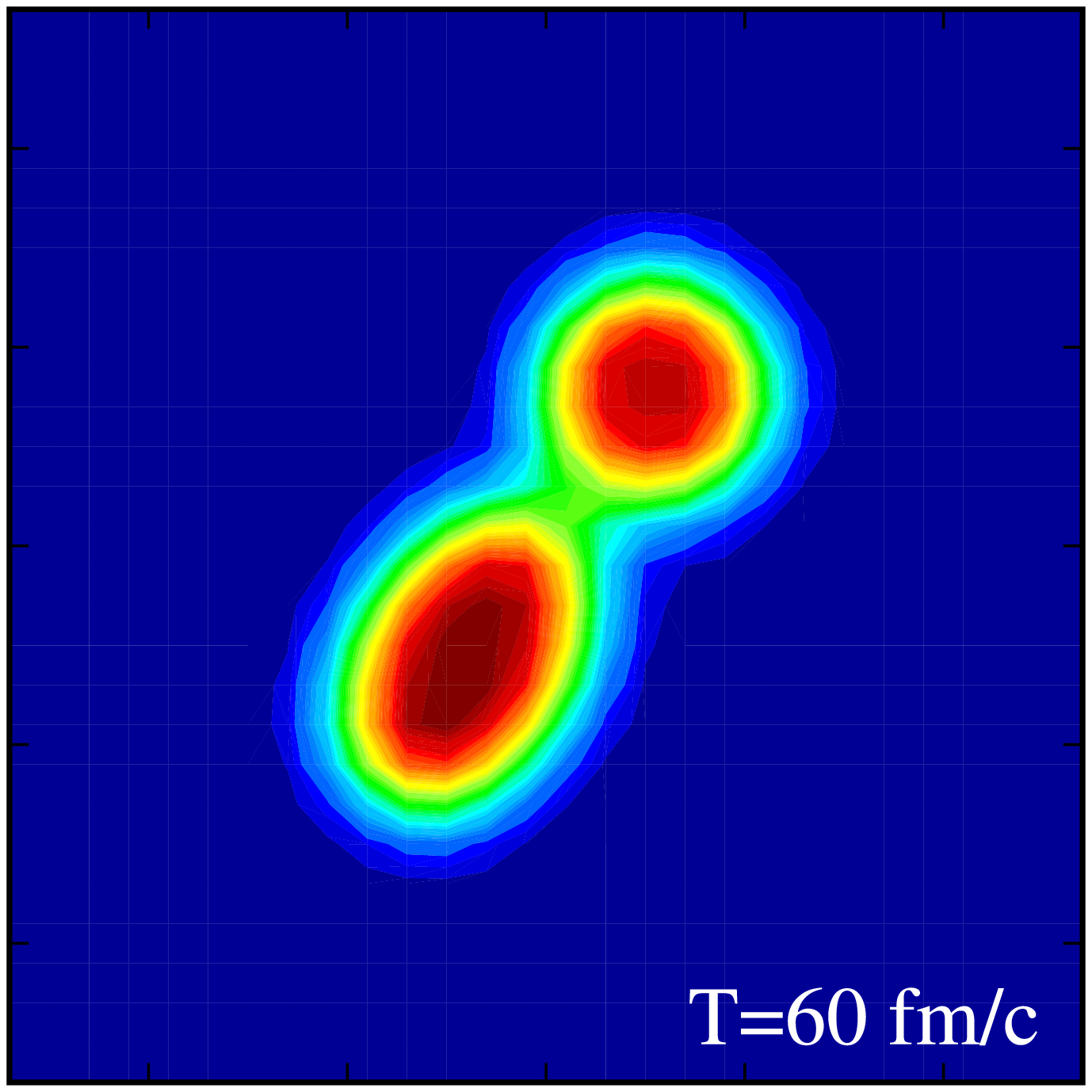}\hspace{-0.05in}
\includegraphics*[scale=0.32]{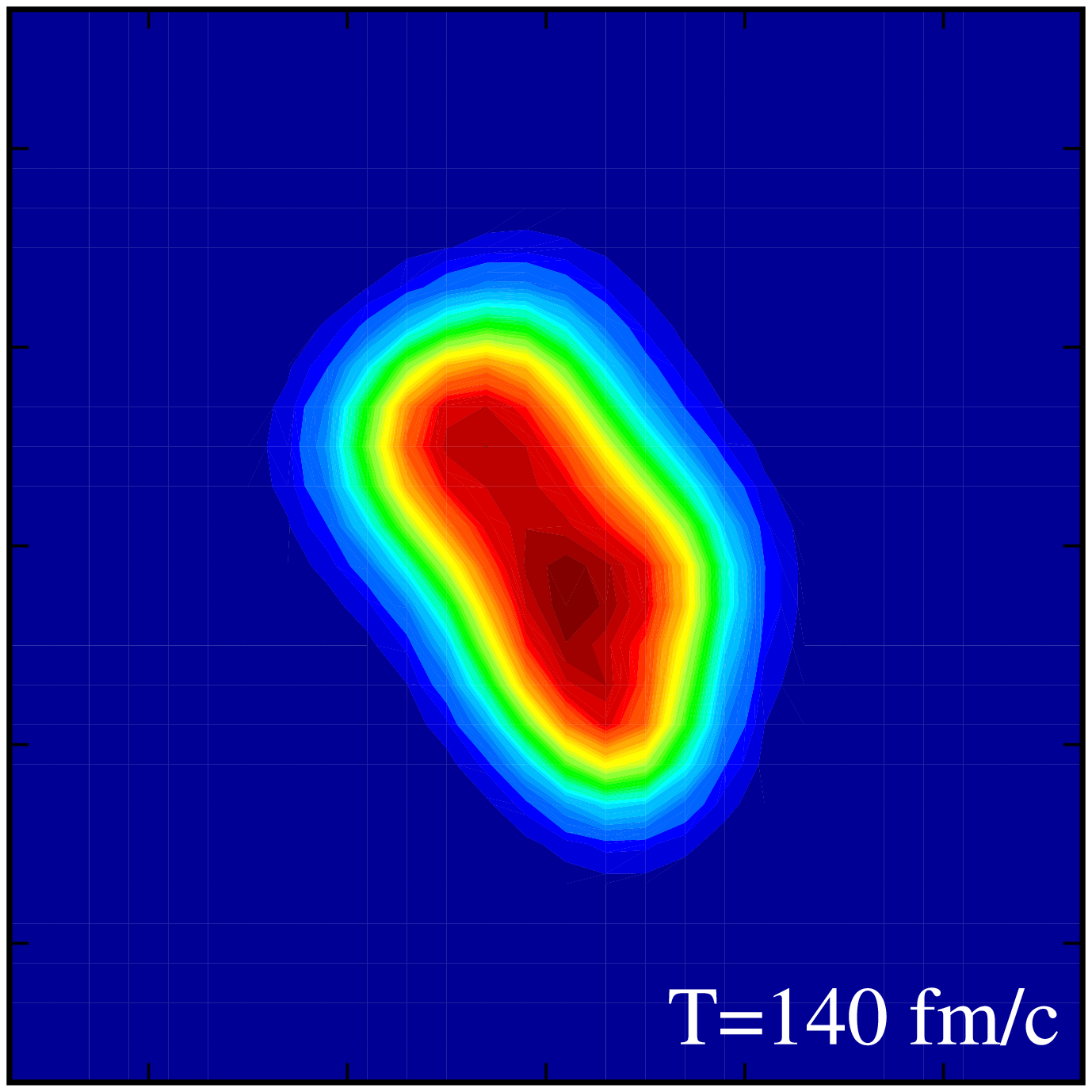}\\ \vspace{-0.05in} \hspace{0.51in}
\includegraphics*[scale=0.32]{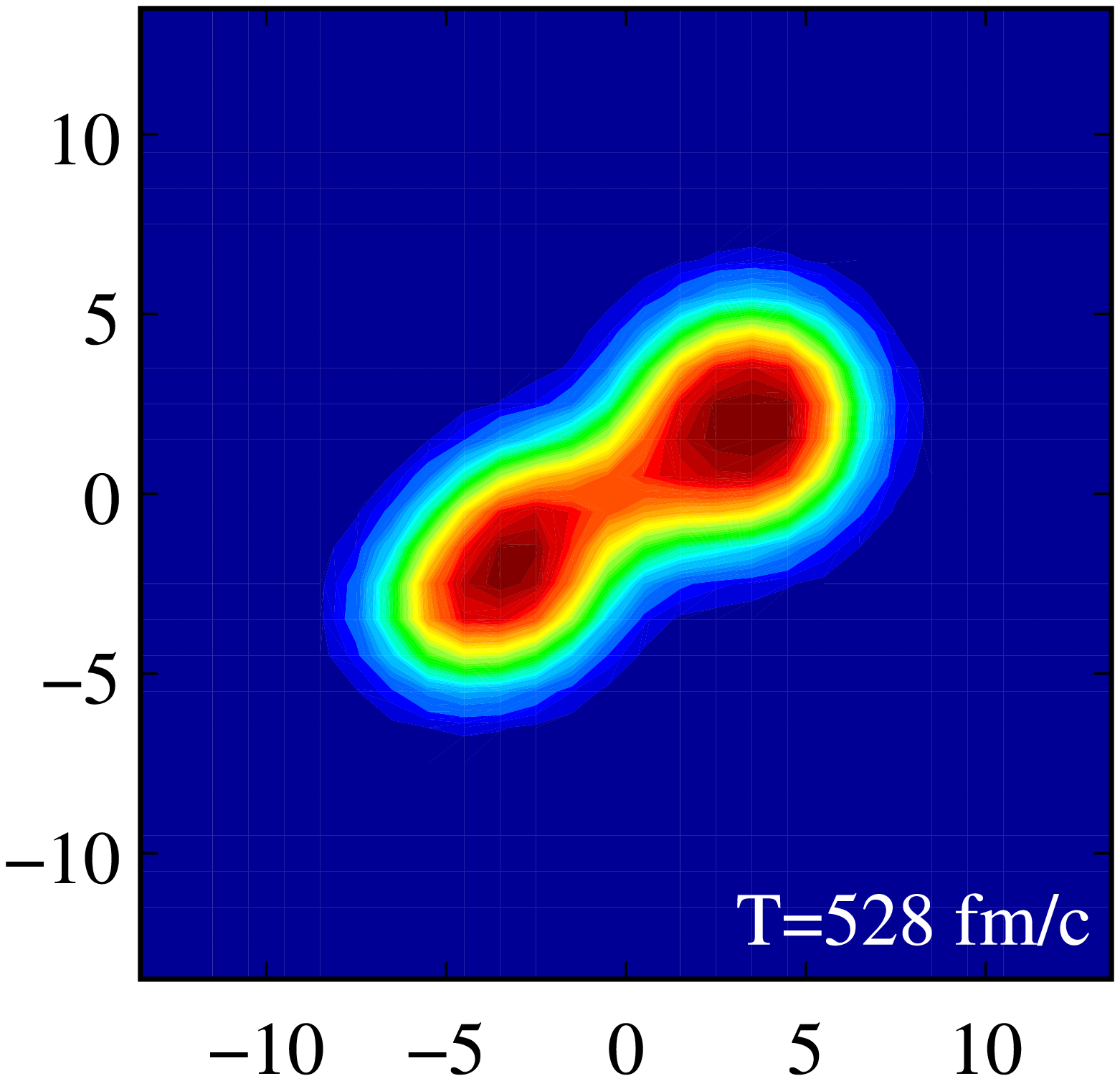}\hspace{-0.05in}
\includegraphics*[scale=0.32]{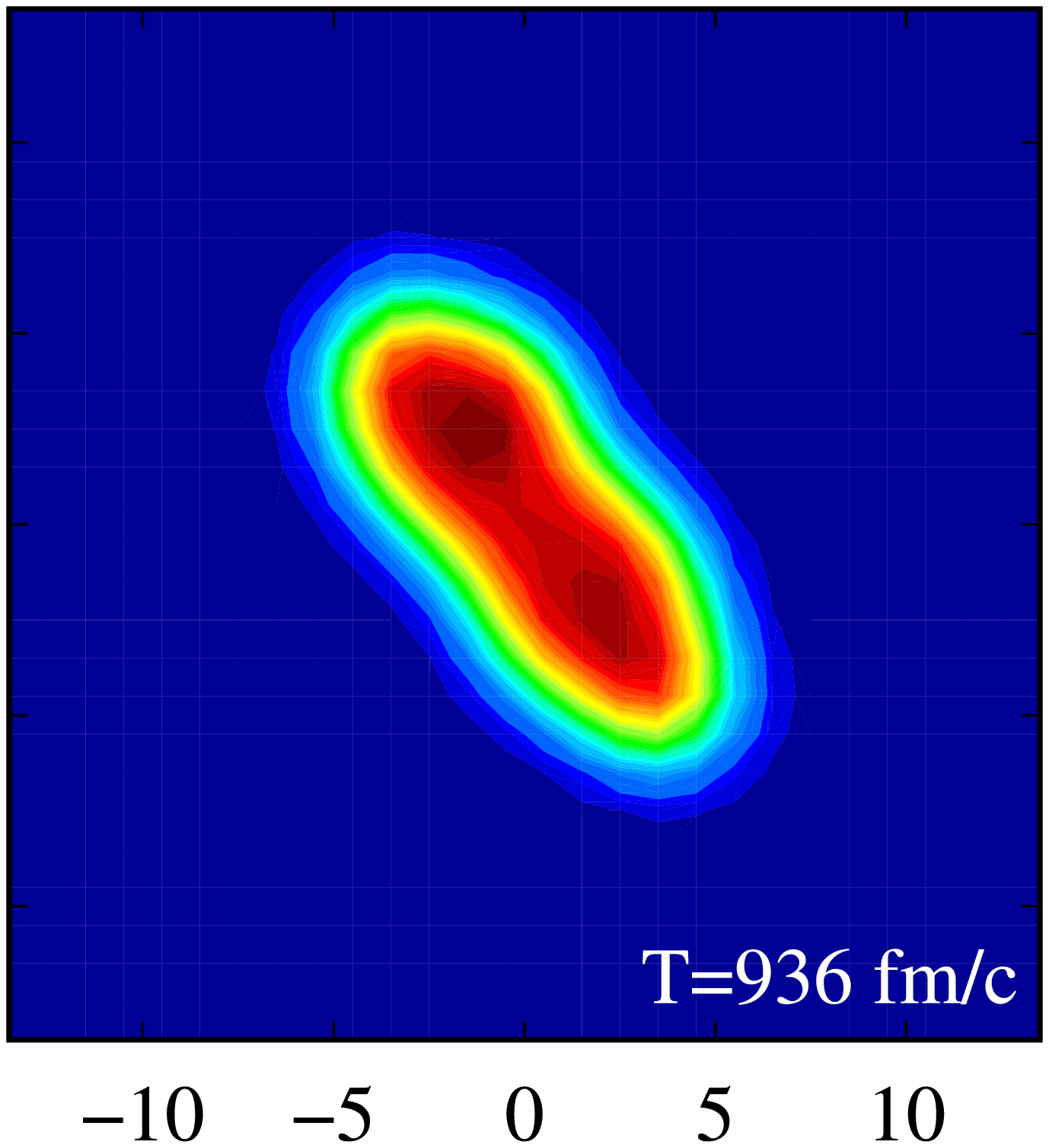}\hspace{-0.05in}
\includegraphics*[scale=0.32]{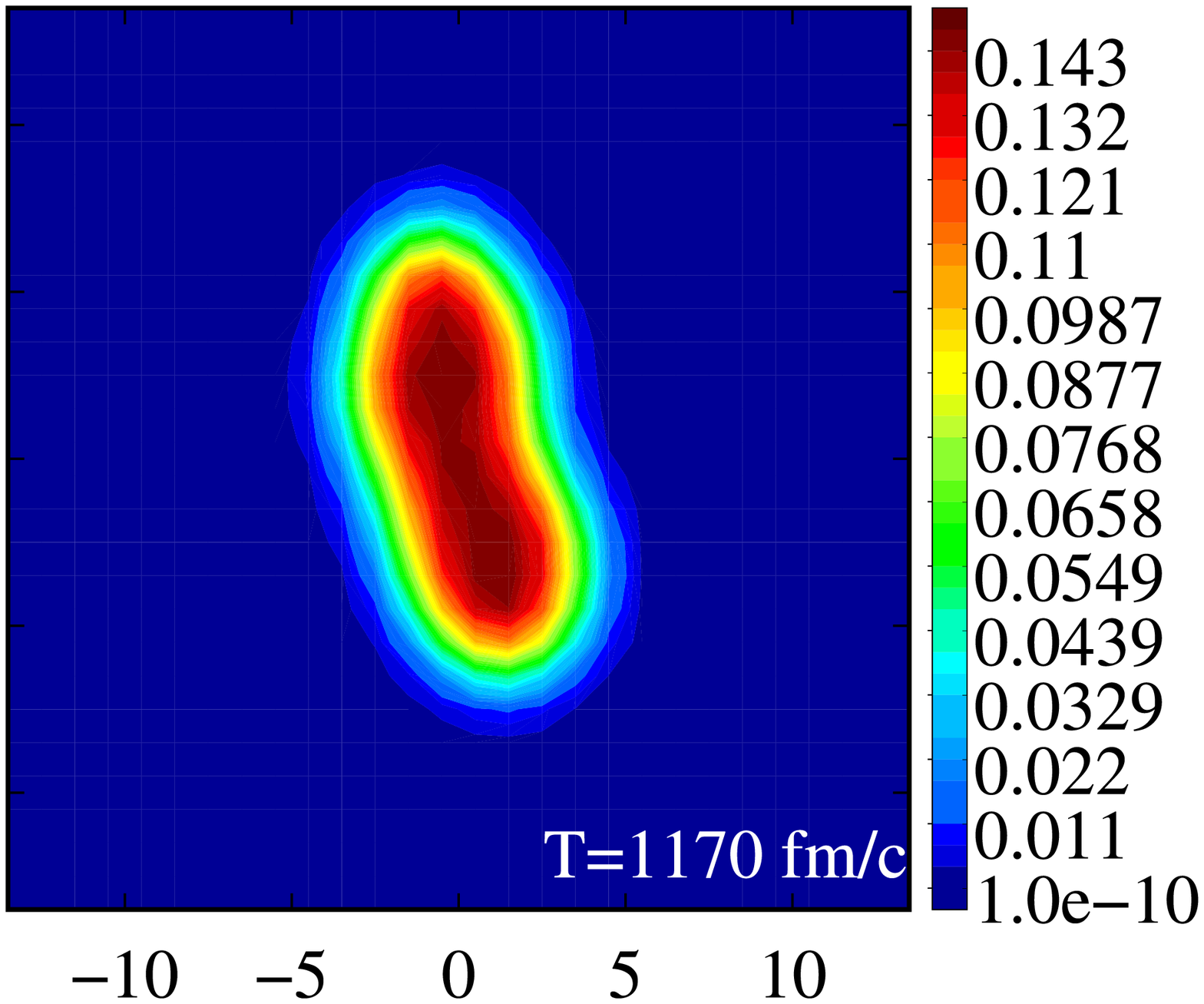}
\caption{\label{fig:time-evol}TDHF time-evolution
for the $^{22}Ne+^{16}O$ collision at an impact parameter of $b=6.35$~fm and initial
Neon orientation angle $\beta=60^{\circ}$ using the SLy4 interaction.
The initial energy is $E_{cm}=95$~MeV. During the evolution the combined system
makes four revolutions.}
\end{center}
\end{figure*}

In order to demonstrate the points made above we have performed TDHF calculations
for the $^{16}O+^{22}Ne$ system at $E_{cm}=95$ MeV. We have chosen this light system
at a relatively
high beam energy so that we can perform the numerical calculations faster.
We have used our new TDHF code~\cite{UO06a}, which works in 3-D and
uses modern Skyrme forces, including the time-odd terms.
For this work we have
used the SLy4 parametrization~\cite{CB98}. In this case Hartree-Fock
calculations result in a spherical $^{16}O$ nucleus and a $^{22}Ne$ nucleus with strong
axial deformation. When no spatial symmetries are assumed Hartree-Fock calculations
generate an orientation for the intrinsic coordinate system, with respect to the
code coordinate system, typically determined by the choice for the initial
single particle states, e.g. harmonic oscillators. Depending on the order in
which the Cartesian oscillator shells are filled we get a particular orientation
for the nucleus. Different orientations with Euler angle $\beta\neq 0$ can
be generated by rotating the coordinate frame in which the initial single
particle states are created with respect to the code frame. The resulting states
will then be oriented with Euler angle $\beta$ with respect to the code frame. A subsequent
rotation perpendicular to this direction would then generate the $\alpha$ rotation.
The Hartree-Fock
iterations in generating the static solutions preserve this orientation since
in an unrestricted geometry all orientations are exactly equivalent.

We have performed calculations by keeping the angle $\alpha$ fixed at $\alpha=0^{\circ}$ and
varying the angle $\beta$ in $10^{\circ}$ intervals. At this energy, all angles up to
$\beta=60^{\circ}$ do not show any fusion for head-on collisions, whereas larger
angles fuse. For each of the orientations leading to fusion we have made a sweep over the impact
parameter to find the maximum impact parameter for fusion. This was initially done
in $1.0$~fm steps and reduced down to $0.05$~fm when the approximate fusion boundary
was located. For large impact parameters the system undergoes multiple revolutions
before fusion or deep-inelastic collision. Starting from an initial separation of
$15$~fm we have followed most of the collisions to
about $1200$~fm/c. Angles smaller than $\beta=60^{\circ}$ contribute to the
deep-inelastic channel. For these orientations a study of the central collisions
show a gradual decrease in the translational kinetic energy between the two final
fragments, ranging from almost $30$~MeV for $\beta=0^{\circ}$ to about $11$~MeV
for $\beta=50^{\circ}$. We also observe particle transfer between the fragments,
in most of the cases the final fragments seem to have exchanged one neutron from
the Neon to the Oxygen. In the fusion regime we have found maximum impact parameters
for fusion as $6.35$~fm, $6.55$~fm, $6.83$~fm, and $6.87$~fm for $\beta$ values
of $60^{\circ}$, $70^{\circ}$, $80^{\circ}$, and $90^{\circ}$, respectively.
In Fig.~\ref{fig:time-evol} we show various time slices of the TDHF evolution
for the case of $\beta=60^{\circ}$ and $b=6.35$~fm.
During this time interval the system makes about four revolutions and eventually
the internal structure shows no memory of the initial structure.

For Euler angles $\beta=60^{\circ}$, $70^{\circ}$,
$80^{\circ}$, and $90^{\circ}$, respectively, we find $^{16}O+^{22}Ne$ fusion cross sections
$d\sigma_{fusion}(E_{cm}=95MeV;\beta,\alpha=0^{\circ}) / sin\beta d\beta d\alpha$
of $633$~mb/sr, $673$~mb/sr, $732$~mb/sr, and $741$~mb/sr.

\section{\label{sec:summary} Conclusions}

To date most TDHF calculations have been limited to the study of collisions
involving spherical systems. Furthermore, approximations made about the
collision geometry (reaction-plane symmetry etc.), in order to make numerical computations
tractable, have lead to the exclusion of deformed systems from such studies.
In this paper, we have presented TDHF calculations involving spherical+deformed
systems without any assumptions regarding collision geometry and using the full
form of the Skyrme interaction. The details of our new TDHF code can be found
in Ref.~\cite{UO06a}. Dealing with deformed nuclei necessitates an approach
to determine the orientation of the deformed system prior to the start of the
TDHF calculations, as such calculations are initialized at relatively small
separations. This alignment is inherently related to multiple E2/E4 excitations
due to the Coulomb interaction between the two nuclei. We have outlined an
approach for calculating this dynamical alignment probability and have shown how
to incorporate it into the cross section calculations. While the alignment
probabilities for various orientations do not vary substantially for light systems
they show considerable preference for particular orientations of heavier systems.
On the other hand, we have also shown that the alignment has a major consequence
for TDHF calculations. We have observed that in the $^{22}Ne+^{16}O$ system
alignments close to the collision axis ($\beta < 60^{\circ}$) results
in no fusion whereas perpendicular alignments lead to fusion. Furthermore, the
alignment naturally affects the impact parameter dependence of fusion for
different orientations. With the advent of computer technology and numerical
methods unrestricted TDHF calculations are becoming more and more feasible.
It is our goal to pursue and improve such calculations in order to provide
a tool for studying heavy-ion collisions with as few computational restrictions
as possible.

\begin{acknowledgments}
This work has been supported by the U.S. Department of Energy under grant No.
DE-FG02-96ER40963 with Vanderbilt University. Some of the numerical calculations
were carried out at the IBM-RS/6000 SP supercomputer of the National Energy Research
Scientific Computing Center which is supported by the Office of Science of the
U.S. Department of Energy.
\end{acknowledgments}


\appendix*

\section{Dynamic alignment of deformed nuclei due to Coulomb excitation}

We present a brief summary of the theory of dynamic alignment of deformed
nuclei during a heavy-ion collision. The alignment results from multiple E2/E4
Coulomb excitation of the ground state rotational band. Details, including a discussion of strong
nuclear interaction effects can be found in Ref.~\cite{O85}. The associated formfactors
for inelastic Coulomb excitation and strong nuclear excitation (proximity potential)
have been derived for collective rotations and surface vibrations in Ref.~\cite{ORS82}.

There exists an extensive literature on the `semiclassical' theory of multiple
Coulomb excitation of heavy ions~\cite{AW75}. In this approach, the excitation
process is described quantum mechanically, and the relative motion of the nuclei is
treated by classical mechanics. The total Hamiltonian consists of the free
Hamiltonian of the target nucleus, $H_0 (X)$, and of the coupling potential
for inelastic Coulomb excitation, $V_C (X,{\bf r}(t))$. The latter depends on the
intrinsic coordinates of the target $(X)$ and on the classical relative trajectory
${\bf r}(t)$.
 
The Coulomb excitation process is determined by the time-dependent Schr\"odinger
equation
\begin{equation}
\left[ H_0 (X) + V_C (X,{\bf r}(t)) \right] \psi (X,t)  = i \hbar
       \frac{\partial}{\partial t} \psi (X,t) \ .
\label{tdse}
\end{equation}
The eigenstates of the target nucleus are given by
\begin{equation}
H_0 (X) \phi_r (X) = E_r \phi_r (X) \ .
\label{h0}
\end{equation}
We expand the time-dependent wavefunction in terms of the stationary eigenstates
$\phi_r$ of the unperturbed Hamiltonian $H_0$
\begin{equation}
\psi (X,t) = \sum_r a_r (t) \phi_r (X) e^{-i E_r t / \hbar}
\label{psiexpan}
\end{equation}
resulting in a system of linear coupled differential equations for
the excitation amplitudes as a function of time
\begin{eqnarray}
i \hbar \dot{a}_r (t) &=& \sum_s a_s (t)  \left \langle \phi_r (X) | V_C (X,{\bf r}(t)) |
    \phi_s (X) \right \rangle        \nonumber \\
 &\times& e^{i (E_r - E_s) t / \hbar} \ .
\label{eq:junk} 
\end{eqnarray}
This system of differential equations is solved numerically by a combination
of the fourth-order Runge-Kutta method and the Adams-Bashforth-Moulton
(predictor-corrector) algorithm~\cite{PT92}. For the classical relative motion
${\bf r}(t)$ we utilize the ``coordinate system B'' defined in Ref.~\cite{AW75},
page 47.

In the specific application of this formalism to dynamic nuclear alignment,
we describe the free Hamiltonian and the corresponding wavefunctions in terms
of the collective rotor model. The degrees of freedom are the three Euler angles $X=(\alpha,\beta,\gamma)$
\begin{equation}
H_0 (X) = T_{rot} (X) \ .
\label{hrot}
\end{equation}
In deformed even-even nuclei, the ground state rotational band has an intrinsic total angular
momentum projection $K=0$; therefore, the collective rotor wavefunctions \cite{BS79} are independent
of the Euler angle $\gamma$ which describes a rotation about the intrinsic symmetry
axis $z'$ (see Fig.~\ref{fig:align_schem})
\begin{eqnarray}
\phi_r (X) &=& \left( \frac{2 J + 1}{8 \pi^2} \right)^{1/2} D^{J\ *}_{M,K=0} (\alpha,\beta,\gamma)  \nonumber \\
&=& \left(2 \pi \right)^{-1/2} Y_{JM} (\beta,\alpha)  \ .
\label{eq:dimk} 
\end{eqnarray}
The probability density at time $t$ to find the nucleus oriented with given Euler angles
$X=(\alpha,\beta)$ is given by $|\psi(X,t)|^2$; by
integration over $\gamma$ we find the corresponding differential orientation probability
\begin{widetext}
\begin{equation}
\frac{dP_{orient}(b,E_{cm},t;\beta,\alpha)}{\sin \beta d \beta d \alpha } = \int_0^{2 \pi} d \gamma
\left| \psi (\alpha, \beta, \gamma; t) \right|^2
 \rightarrow \left| \sum_{J,M} a_{JM} (t) Y_{JM} (\beta,\alpha) e^{-i E_J t / \hbar} \right|^2 \ .
\label{eq:orient_prob} 
\end{equation}
\end{widetext}


\begin{thebibliography}{99}

\bibitem{Ba80} R. Bass, {\it Nuclear Reactions with Heavy Ions}, (Springer-Verlag, New York, 1980).

\bibitem{Li03} J. F. Liang et al., Phys. Rev. Lett. {\bf 91}, 152701 (2003);
               Erratum: Phys. Rev. Lett. {\bf 96}, 029903 (2006).

\bibitem{Ho02} S. Hofmann et al., Eur. Phys. J. A {\bf 14}, 147 (2002).

\bibitem{Gi03} T. N. Ginter et al., Phys. Rev. C {\bf 67}, 064609 (2003).

\bibitem{Og04} Yu. Ts. Oganessian et al., Phys. Rev. C {\bf 69}, 021601(R) (2004).

\bibitem{Mo04} K. Morita et al., Eur. Phys. J. A {\bf 21}, 257 (2004).

\bibitem{II05} T. Ichikawa, A. Iwamoto, P. M\"oller, and A.J. Sierk, Phys. Rev. C {\bf 71}, 044608 (2005).

\bibitem{SE78} R. G. Stokstad, Y. Eisen, S. Kaplanis, D. Pelte, U. Smilansky, and
               I. Tserruya, Phys. Rev. Lett. {\bf 41}, 465 (1978).

\bibitem{KH02} K. Nishio, et al., J. Nucl. Radiochemical Sci., {\bf 3}, 89 (2002).

\bibitem{FG04} G. Fazio et al., Eur. Phys. J. A {\bf 19}, 89 (2004).

\bibitem{SC04} C. Simenel, Ph. Chomaz, and G. de France, Phys. Rev. Lett. {\bf 93}, 102701-1 (2004).

\bibitem{TB84} N. Takigawa and G.F. Bertsch, Phys. Rev. C {\bf 29}, 2358 (1984).

\bibitem{BT98} A. B. Balantekin and N. Takigawa, Rev. Mod. Phys. {\bf 70}, 77 (1998).

\bibitem{RO83} M. J. Rhoades-Brown and V. E. Oberacker, Phys. Rev. Lett. {\bf 50}, 1435 (1983).

\bibitem{LP84} S. Landowne and S. C. Pieper, Phys. Rev. C {\bf 29}, 1352 (1984).

\bibitem{RP84} M. J. Rhoades-Brown and M. Prakash, Phys. Rev. Lett. {\bf 53}, 333 (1984).

\bibitem{Esb04} H. Esbensen, Prog. Theor. Phys. Suppl. {\bf 154}, 11 (2004).

\bibitem{HR99} K. Hagino, N. Rowley, and A.T. Kruppa, Comp. Phys. Comm. {\bf 123}, 143 (1999).

\bibitem{JN82} J. W. Negele, Rev. Mod. Phys. {\bf 54}, 913 (1982).

\bibitem{BFH85}  P. Bonche, H. Flocard, P.-H. Heenen, S.J. Krieger, and M.S. Weiss, Nucl.Phys. {\bf A443}, 39 (1985).

\bibitem{U91} A. S. Umar, M. R. Strayer, J.-S. Wu, D. J. Dean, and M. C. G\"u\c cl\"u,
              Phys. Rev. C {\bf 44}, 2512 (1991).

\bibitem{UO05} A. S. Umar and V. E. Oberacker, Eur. Phys. J. A {\bf 25}, s01, 553 (2005).

\bibitem{UO06a} A. S. Umar and V. E. Oberacker, Phys. Rev. C (submitted); http://arxiv.org/abs/nucl-th/0603038.

\bibitem{ENSDF} Evaluated Nuclear Structure Data File (ENSDF), National Nuclear Data Center,
                Brookhaven National Laboratory, http://www.nndc.bnl.gov/ensdf/.

\bibitem{TOI} LBNL Isotopes Project, http://ie.lbl.gov/toi.htm.

\bibitem{BS79} D. M. Brink and G.R. Satchler, {\it Angular Momentum}, 2nd ed., (Clarendon Press, Oxford, 1979).

\bibitem{CB98} E. Chabanat, P. Bonche, P. Haensel, J. Meyer, and R. Schaeffer,
               Nucl. Phys. {\bf A635}, 231 (1998); Nucl. Phys. {\bf A643}, 441 (1998).

\bibitem{O85} V. E. Oberacker, Phys. Rev. C {\bf 32}, 1793 (1985).

\bibitem{ORS82} V. E. Oberacker, M. J. Rhoades-Brown, and G. R. Satchler, Phys. Rev. C {\bf 26}, 129 (1982).

\bibitem{AW75} K. Alder and A. Winther, {\it Electromagnetic Excitation}, (North Holland, Amsterdam, 1975).

\bibitem{PT92} W. H. Press, S. A. Teukolsky, W. T. Vetterling, and B. P. Flannery,
               {\it Numerical Recipes},  2nd ed., (Cambridge University Press,  Cambridge, 1992).

\end{thebibliography}


\end{document}